\documentclass[10pt,conference]{IEEEtran}
\usepackage{alltt                                    
          , multirow
          , booktabs
          , listings
          , graphicx
          ,float
	,cite
          ,verbatim
         ,mathtools
	,url
	,amsmath
}
\usepackage[table]{xcolor}
\usepackage[numbers]{natbib}     
\usepackage{syntax}
\usepackage{algorithmic, algorithm}
\usepackage{enumitem}
\usepackage{threeparttable}
\usepackage[T1]{fontenc}

\usepackage{expl3}
\ExplSyntaxOn
\newcommand\latinabbrev[1]{
  \peek_meaning:NTF . {
    #1\@}%
  { \peek_catcode:NTF a {
      #1., \@ }%
    {#1., \@}}}
\ExplSyntaxOff

\newcommand{\CASE}[1]{\STATE \textbf{case} #1\textbf{:} \begin{ALC@g}}
\newcommand{\ENDCASE}{\end{ALC@g}}

\newcommand{\DEFAULT}{\STATE \textbf{default:} \begin{ALC@g}}
\newcommand{\ENDDEFAULT}{\end{ALC@g}}
\newcommand{\DEFAULTLINE}[1]{\STATE \textbf{default:} }

\newsavebox{\supbox}
\newcommand{\bsup}{\begin{lrbox}{\supbox}$\tt\scriptstyle}
\newcommand{\esup}{$\end{lrbox}{}^{\usebox{\supbox}}}
\def\eg{\latinabbrev{e.g}}
\def\ie{\latinabbrev{i.e}}

\definecolor{lightpurple}{rgb}{0.8,0.8,1}
\definecolor{codebg}{RGB}{255,255,255}
\definecolor{commentcolor}{RGB}{11,140,11}
\definecolor {red}{rgb}{255,0,0}
\begin{document}
%

\title{RACK: Code Search in the IDE using Crowdsourced Knowledge \vspace{-0.3cm}}

%
%
%

\author{\IEEEauthorblockN{Mohammad Masudur Rahman  ~~~ Chanchal K. Roy ~~~ David Lo$^\dagger$}
\IEEEauthorblockA{University of Saskatchewan, Canada, $^\dagger$Singapore Management University, Singapore\\
\{masud.rahman, chanchal.roy\}@usask.ca, davidlo@smu.edu.sg}
}



%
\maketitle
\begin{abstract}
Traditional code search engines often do not perform well with natural language queries since they mostly apply keyword matching.
These engines thus require carefully designed queries containing information about programming APIs for code search. 
Unfortunately, existing studies suggest that preparing an effective query for code search is both challenging and time consuming for the developers.
In this paper, we propose a novel code search tool--RACK--that returns relevant source code for a given code search query written in natural language text.  
The tool first translates the query into a list of relevant API classes by mining keyword-API associations from the crowdsourced knowledge of Stack Overflow, and then applies the reformulated query to GitHub code search API for collecting relevant results.
Once a query related to a programming task is submitted, the tool automatically mines relevant code snippets from thousands of open-source projects, and displays them as a ranked list within the context of the developer's programming environment--the IDE.
Tool page: http://www.usask.ca/$\sim$masud.rahman/rack 
\end{abstract}



\begin{IEEEkeywords}
Code search, query reformulation, keyword-API association, crowdsourced knowledge, Stack Overflow
\end{IEEEkeywords}

\IEEEpeerreviewmaketitle


\section{Introduction}
Studies show that software developers on average spend about 19\% of their development time in web search where they mostly look for relevant code snippets for their programming tasks \cite{twostudy}. 
Code search engines--\emph{Open Hub, Koders, GitHub search} and \emph{Krugle}--index thousands of open source projects which are a potential source for such snippets \cite{portfolio}.
Unfortunately, preparing an effective query for code search containing information about relevant APIs is not only a challenging task but also is time-consuming for the developers \cite{kevic,twostudy}.
Previous study also reported that on average, developers performed poorly in coming up with good search terms regardless of their experience levels \cite{kevic}.
Thus, a tool that automatically translates a natural language query from the developer into a set of relevant API classes or methods (\ie\ search-engine friendly query) and then returns relevant source code snippets, can greatly assist the developers in their tasks.
Our paper addresses this research problem, and provides automatic tool support both in preparing search queries and in performing code search conveniently.

\begin{figure*}[!t]
\centering
\includegraphics[width=7in ]{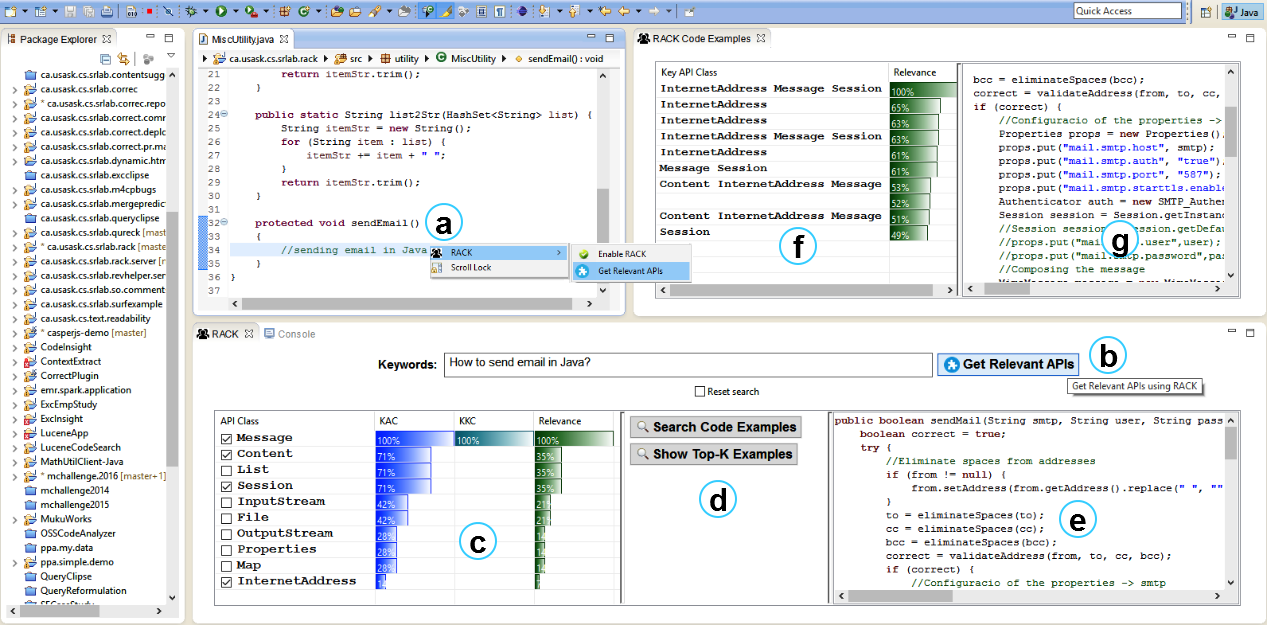}
\vspace{-.3cm}
\caption{User Interface of RACK}
\label{fig:ide}
\vspace{-.5cm}
\end{figure*}

Existing studies accept one or more natural language queries, and return relevant API classes and methods by analyzing
feature request history and API documentations \cite{feature}, API invocation graphs \cite{conngraph}, library usage patterns, code surfing behaviour of the developers and API invocation chains \cite{portfolio}. 
Although these techniques perform well in different working contexts, they share a set of limitations and fall short to address our research problem.
First, each of these techniques \cite{portfolio,conngraph,feature} exploits textual similarity measure (\eg\ Dice's coefficients \cite{conngraph}) for candidate API selection. This warrants that the search query should be carefully prepared, and it should contain keywords similar to the API names. 
In other words, the developer should possess a certain level of experience on the target APIs to actually use those techniques. 
Second, API names and search queries are generally provided by different developers who may use different vocabularies to convey the same concept. 
Concept/feature/concern location community have termed it as \emph{vocabulary mismatch problem} \cite{qeffect}. Textual similarity based techniques often suffer from this problem. 
Hence, the performance of these techniques is not only limited but also subject to the identifier naming practices adopted in the codebase under study. 


In this paper, we propose a novel code search tool--\emph{RACK}--that accepts an unstructured natural language query (\ie\ does not require API information) from a developer as input and returns relevant code snippets as output from thousands of open source projects.
The tool first captures the developer's intent for code search from two working contexts (\eg\ code comments) within the IDE as a query, translates the query into relevant API classes automatically, and then collects the relevant code examples from GitHub search API by applying them. 
While each question in Stack Overflow Q \& A site summarizes a programming problem/task, the corresponding answers often suggest appropriate APIs that solve the problem. We thus mine thousands of questions and corresponding accepted answers from Stack Overflow, and translate the natural language query (\ie\ programming task) into relevant API classes by exploiting the keyword-API associations from Stack Overflow. Thus, the tool works both as a query recommender and as a code search engine. We package our solution as an Eclipse IDE plug-in that allows the developers to perform code search within the IDE, and thus, they can avoid the annoying context-switching issue. To summarize, our tool provides the following features to support the developers in code search:
\begin{enumerate}[topsep=.2ex]
\item the proposed tool automatically translates an unstructured natural language query referring to a programming task into relevant API classes for the task. 
\item determines relevance of the returned API classes based on keyword-API associations mined from thousands of programming questions and solutions of Stack Overflow.
\item mitigates the vocabulary mismatch problem faced by existing techniques and traditional code search engines.
\item integrates GitHub search API into the IDE for IDE-based code search and convenient result display.
\item offers meaningful relevance insights for both search queries and search results unlike any traditional search.

\end{enumerate}

While this paper focuses on tool aspect of our approach, we refer the readers to the original paper \cite{saner2016masud} for further details.


\vspace{-.1cm}
\section{RACK}\label{sec:rack}
Fig. \ref{fig:ide} shows the user interface of \emph{RACK}, where we contribute in (b)--(c) query suggestion panel, (d)--(e) code search panel, and (f)--(g) result panel of the interface. This section discusses different technical features provided by our tool.

\textbf{(1) Automatic Suggestion of Code Search Queries}: 
RACK automatically suggests relevant keywords (\ie\ API classes) for code search  given that preparing an effective query for a programming task
 is a significant challenge \cite{kevic,sitir}. 
Our tool overcomes this challenge by mining thousands of programming problems and their corresponding solutions from Stack Overflow Q \& A site. 
It captures a developer's intent for code search from various working contexts within the IDE, and suggests a list of appropriate keywords for code search with meaningful insights (\ie\ relevance scores).

\textbf{(i) Working Contexts}: RACK captures natural language queries (\ie\ developer's intents) for code search from two working contexts of a developer-- \emph{source code comment}  and \emph{traditional search box}.
Once the developer intends to accomplish a programming task by stating in the header comment of a method, our tool captures the comment as an initial query for reformulation (\eg\ Fig. \ref{fig:ide}-(a)). In the second case, the developer provides an initial query written using unstructured natural language texts, and RACK captures the query from the traditional search box (Fig. \ref{fig:ide}-(b)) for relevant API suggestion.        


\textbf{(ii) Mining of Relevant API Classes:} In Stack Overflow, users often submit questions focusing programming tasks (\eg\ ``How can I generate MD5 hash?"), and the corresponding answers suggest relevant APIs (\eg\ \texttt{MessageDigest}) for accomplishing those tasks. RACK accesses a database of 344K such questions and answers, learns keyword--API associations, and then suggests relevant API classes for a given task.    

\textbf{(iii) API Suggestion and Query Reformulation}: Once the natural language query (\ie\ initial query) is submitted, RACK suggests the Top-10 relevant API classes for the task in the query (Fig. \ref{fig:ide}-(c)). Not only the suggestions are provided as a ranked list but also our tool explains why a particular API is relevant by visualizing three meaningful scores-- Keyword--API Co-occurrence (KAC), Keyword--Keyword Coherence (KKC), and Overall Relevance \cite{saner2016masud}. Being equipped with such ranking and insights, a developer can easily choose appropriate APIs by marking them checked and initiate the code search.   


\textbf{(2) IDE-Based Code Search}: RACK not only provides an IDE-based code search feature but also assists the developer in result analysis with a customized view. It provides two flexible code search options and displays the results with meaningful insights (\ie\ relevance scores) within the IDE.  

\textbf{(i) Code Search Options and Backend:} RACK provides two options--Top-1 search and Top-K search--for performing code search in the IDE (Fig. \ref{fig:ide}-(d)). Once the relevant API classes are suggested (by the tool) and appropriate classes (\ie\ search keywords) are chosen by the developer, the tool returns the topmost relevant code snippet from thousands of open source projects of four large organizations--\emph{Apache, Eclipse, Google} and \emph{Facebook}. We integrate GitHub code search API in the backend for collecting the relevant source code files from which the most relevant method body is extracted using Abstract Syntax Tree (AST) parsing and textual similarity analysis with the query.
In the second case, RACK returns the Top-K (\eg\ $K=10$) code snippets based on their relevance for further analysis by the developer. One can also reset the whole process by checking the check box provided by the tool.

\textbf{(ii) Mitigation of Vocabulary Mismatch Issue:} Textual similarity based search techniques (\eg\ Vector Space Model) generally suffer from this issue when unstructured natural language queries are used for code search \cite{qeffect,vocaprob}. Since RACK translates the initial search query into relevant API classes that come from standard libraries or development toolkits (\ie\ from a single vocabulary), such issue is mitigated.

\textbf{(iii) Result Display and Insights:} RACK not only shows the code search results as a meaningful ranked list (\ie\ with relevance insights) but also adds an in-line source code viewer for detailed analysis of the results (Fig. \ref{fig:ide}-(e)--(g)). Each result from the list is annotated using the matched keywords from the query which provides additional intuition about its relevance. The code viewer is enabled with \emph{syntax highlighting} which ensures a convenient analysis of the code by the developer.

\textbf{(3) Performance Optimization}: While the query reformulation step requires relational database access, the code search step involves GitHub API access and significant static analysis of the source code. In both steps, RACK applies Java multi-threading for optimized computation and response time. To date, our reformulation takes $\le$10 seconds and the search takes $\le$2 seconds on average which are close to real time.

\textbf{(4) Seamless Integration and Dynamic Corpus:} RACK adopts a client-server architecture where the Eclipse IDE plug-in is the client module, and the server module (\ie\ query reformulation engine) is hosted as a web service. That is, any tool capable of making HTTP calls can consume our query reformulation service, which demonstrates RACK's modularity. On the other hand, the use of GitHub API ensures that RACK always returns relevant code from an automatically evolving and carefully indexed large source code corpus.

\begin{table}
\centering
\caption{Suggested API Classes for the Use Case}\label{table:suggested}
\vspace{-.2cm}
\resizebox{3.5in}{!}{%
\begin{threeparttable}
\begin{tabular}{l|c|c|c||l|c|c|c}
\hline
\textbf{API} & \textbf{KAC} & \textbf{KKC} & \textbf{Relevance} & \textbf{API} & \textbf{KAC} & \textbf{KKC} & \textbf{Relevance}\\
\hline
\hline
\texttt{File} & 0.60 &1.00 &1.00 & \texttt{Element} & 0.60 & 0.46  & 0.66\\
\hline
\texttt{Document} & 1.00 & 0.46 &0.91 & \texttt{Jsoup}& 0.40 &0.00 & 0.25\\
\hline
\texttt{List} & 0.90 &0.22 & 0.70 & \texttt{Elements}& 0.20 &0.00 & 0.19 \\
\hline
\end{tabular}
\centering
\end{threeparttable}
}
\vspace{-.6cm}
\end{table}

\section{A Use Case Scenario}\label{sec:usecase}
By means of a use case scenario, we attempt to explain how \emph{RACK} can help a software developer in accomplishing a programming task  within the IDE.


Suppose a developer, Alice, is attempting to develop a Java application that parses an HTML page (\eg\ Yahoo! finance page), and extracts certain items of her interest (\eg\ stock price). 
However, she lacks necessary experience and thus is looking for a working code example that performs the same or similar task.
She formulates a query--\emph{``parsing html in Java"}, and submits to a web search engine (\eg\ Google). The search engine leads her to a list of programming Q \& A pages and API documentations.
Now, she needs to go through the pages carefully containing a large body of texts. While these pages might be useful for improving her knowledge on parsing, choosing relevant code examples from them is not only a time consuming but also a non-trivial job. 
She also submits the same natural language query to a code search engine (\eg\ GitHub), but the returned results were not promising. In short, she (1) fails to collect a succinct and working code example comfortably from the web search results due to the noise in the content, (2) does not get a relevant result from the code search engine due to its inherent limitation--vocabulary mismatch issue between the query and source code, (3) finds the display of neither web search results nor code search results helpful for post-search analysis (\ie\ trying out examples).

\begin{figure}[!t]
\centering
\includegraphics[width=3.5in ]{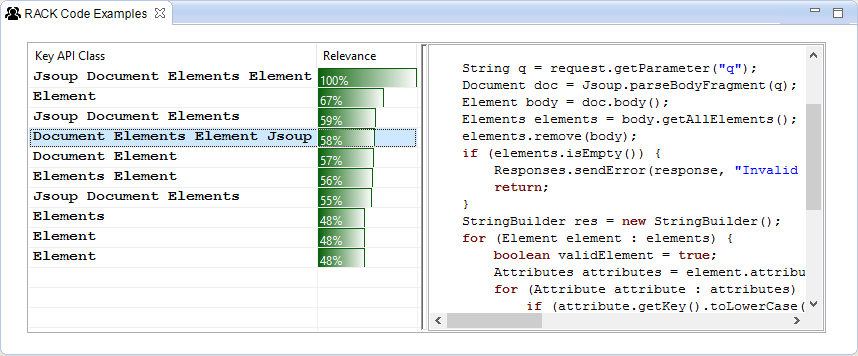}
\vspace{-.7cm}
\caption{Top-10 code search results for the use case}
\label{fig:cmsearch}
\vspace{-.5cm}
\end{figure}
         
Now, let us assume that Alice has installed \emph{RACK} in her IDE, and she encounters the same programming challenge. Our tool captures her natural language query from the code comment (\eg\ Fig. \ref{fig:ide}-(a)), and automatically suggests a ranked list of relevant API classes along with three relevance insights (\ie\ KAC, KKC and Relevance) for the task.
Table \ref{table:suggested} shows the Top-6 APIs suggested by RACK. Among them four (\ie\ 67\%) classes--\texttt{Document, Element, Jsoup} and \texttt{Elements}--are related to HTML parsing.
She can play along with the top API classes, reformulate the initial query, and instantly try out the working code examples returned by the reformulated query.
Existing study reported that developers frequently experiment with and learn from working code examples \cite{twostudy}.
Fig. \ref{fig:cmsearch} shows the Top-10 relevant code snippets returned by RACK for this use case which are mined
from thousands of open source projects using GitHub code search API.
Not only our tool provides the relevance estimate for each result but also it annotates them with matched keywords and adds an in-line source code viewer.
Such information and feature are likely to assist one in analyzing the code results  more conveniently.

Thus, RACK (1) provides Alice one or more succinct and relevant code example(s) without much effort or time spent (\ie\ 10-15 seconds), (2) overcomes the  vocabulary mismatch issue of a traditional code search engine through effective query reformulation (\ie\ relevant API classes), and (3) displays the results with meaningful insights and convenient viewing panel. In short, our tool does all the heavy lifting on behalf of Alice and provides a better alternative than the traditional means for code search and her problem solving.

\begin{figure}[!t]
\centering
\includegraphics[width=3.5in ]{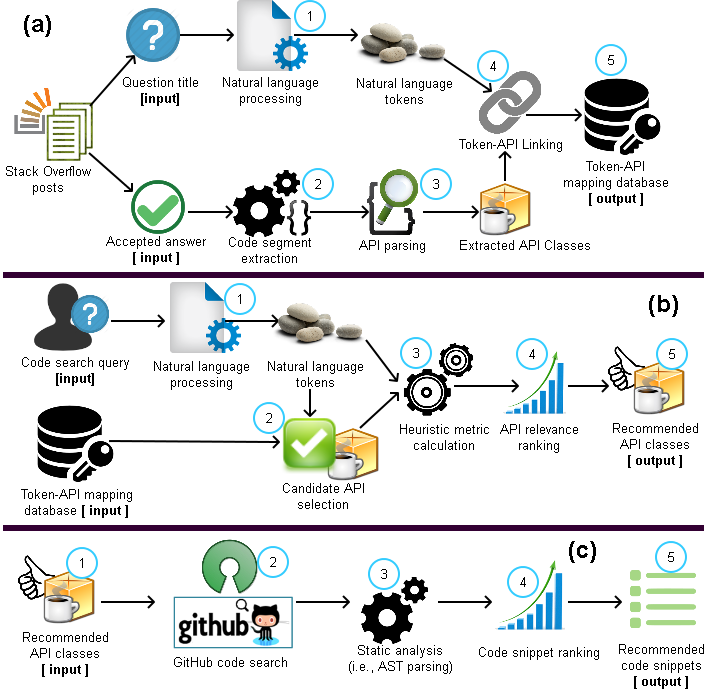}
\vspace{-.8cm}
\caption{Schematic diagram of RACK -- (a) Construction of keyword-API mapping database, (a) Reformulation of a natural language query, and (c) Code snippet search}
\label{fig:sysdiag}
\vspace{-.5cm}
\end{figure}

\section{Working Methodology}\label{sec:methodology}
Fig. \ref{fig:sysdiag} shows the schematic diagram of our proposed tool.
This section discusses the internal structures and working methodologies of the tool in brief, while we refer the readers to the original paper \cite{saner2016masud} for details. 

\textbf{Construction of Keyword-API Mapping Database:} 
We first construct our keyword-API mapping database by carefully analyzing 344K programming questions and corresponding accepted answers (\ie\ solutions) from Stack Overflow Q \& A site. The keywords are collected from the question titles using natural language preprocessing whereas 
the API classes are extracted from the answers through island parsing (\ie\  Steps 1--3, Fig. \ref{fig:sysdiag}-(a)). Then we capture the inherent associations between the keywords and the API classes from each question-answer pair, and construct the keyword-API mapping database (Steps 4-5, Fig. \ref{fig:sysdiag}-(a)).
RACK accesses this database for reformulating a natural language query.

\textbf{Query Reformulation}:  Once an initial query written in unstructured natural language is submitted to RACK, the query is sanitized through natural language preprocessing (\ie\ stop word removal, token splitting, stemming) and converted into a vector of keywords. Then those keywords are used to collect the candidate API classes from the mapping database using two heuristics--KAC and KKC (\ie\ Steps 1--3, Fig. \ref{fig:sysdiag}-(b)). Then the candidates are ranked based on their likelihood (\ie\ derived from KAC) and coherence (\ie\ derived from KKC) with the keywords. Finally, the tool returns a ranked list of relevant API classes (with relevance estimates) as a reformulation of the initial query (\ie\ Steps 4, 5, Fig. \ref{fig:sysdiag}-(b)).

\textbf{Code Search in the IDE}:
Once a reformulated query containing appropriate/relevant API classes is submitted to RACK, it uses GitHub search API and collects relevant source code files from thousands of open source projects hosted by four large organizations--Apache, Eclipse, Google and Facebook. Given that developers are often interested in trying out the code snippets performing a particular task, we parse all the methods from each source file using AST-based parsing (\eg\ \emph{Javaparser} library) (\ie\ Steps 1--3, Fig. \ref{fig:sysdiag}-(c)). GitHub API returns a relevance score for each result file which we combine with the textual similarity scores (with the search query) of all the methods extracted from that file. This combination provides a combined relevance for each code snippet (\ie\ method), and RACK finally returns a ranked list of relevant code snippets within the IDE (\ie\ Steps 4, 5, Fig. \ref{fig:sysdiag}-(c)).

\section{Performance}\label{sec:performance}
Since our original paper \cite{saner2016masud} claims main contributions in the API recommendation for query reformulation, that part of RACK was rigorously evaluated and validated. 
To evaluate the API suggestion performance, we conduct experiments using 150 code search queries randomly chosen from three programming tutorial sites--\emph{KodeJava, Java2s} and \emph{JavaDB}.
The evaluation shows that RACK was able to suggest at least one relevant API class for 79\% of the queries within the Top-10 API suggestions, which is highly promising according to the literature. Comparison with the state-of-the-art--\citet{feature}--not only validated our performance but also confirmed the superiority of RACK in relevant API suggestion. Since our reformulated queries contain gold set API classes and we exploit GitHub API for code search, our queries are also  likely to return relevant code snippets given that GitHub applies keyword matching in source code search. 

\section{Conclusion \& Future Work}\label{sec:conclusion}
To summarize, we propose a novel IDE-based code search tool--\emph{RACK}--that returns relevant code snippets for natural language queries unlike the traditional code search engines. It exploits crowdsourced knowledge from Stack Overflow for query reformulation (details in the original paper \cite{saner2016masud}), and then applies the reformulated queries to collecting relevant code from GitHub search API. 
In future, we plan to conduct an exhausted user study with the tool involving prospective participants.
We also plan to apply the inherent mapping between keywords and API classes mined from Stack Overflow posts to several other software maintenance activities such as concept location, bug localization and source code re-documentation.

\section*{Acknowledgement}
This research was supported in part by the Natural Sciences and Engineering Research Council of Canada (NSERC) and the Singapore Ministry of Education (MOE) Academic Research Fund (AcRF) Tier 1 grant.

\bibliographystyle{plainnat}
\scriptsize
\bibliography{sigproc}  
%
%

\end{document}